\documentclass[%
reprint,
superscriptaddress,
aps,
prb,
]{revtex4-2}

\usepackage{amsmath}
\usepackage{amssymb}
\usepackage{multirow}
\usepackage{float}
\usepackage[pdftex]{graphicx}
\usepackage{xspace}
\usepackage{dsfont}
\usepackage{bbold}
\usepackage{dcolumn}
\usepackage{bm}
\usepackage{todonotes}
\usepackage[colorlinks=true,citecolor=blue,linkcolor=blue,linktocpage=true,pagebackref=false]{hyperref}
\hypersetup{colorlinks=true,citecolor=blue,linkcolor=blue,filecolor=blue,urlcolor=blue}

\usepackage{comment}
\usepackage[normalem]{ulem}

\newcommand{\ou}{%
  \mathrel{%
    \vcenter{\offinterlineskip
      \ialign{##\cr$<$\cr\noalign{\kern-1.5pt}$>$\cr}%
    }%
  }%
}

\usepackage{xcolor}           
\usepackage[normalem]{ulem}   

\begin{document}


\title{Anomalous Localization in Magnetically Doped Two-Dimensional Topological Insulators}

\author{Felippe Amorim}
\thanks{These authors contributed equally to this work}
\affiliation{Instituto de Física Teórica, S\~ao Paulo State University, S\~ao Paulo, Brazil}
\affiliation{School of Physics, Trinity College Dublin, Dublin 2, Ireland}

\author{Washington F. dos Santos}
\thanks{These authors contributed equally to this work}
\affiliation{Instituto de Física Teórica, S\~ao Paulo State University, S\~ao Paulo, Brazil}

\author{Mauro~S.~Ferreira}
\affiliation{School of Physics, Trinity College Dublin, Dublin 2, Ireland}
\affiliation{Advanced Materials and Bioengineering Research (AMBER) Centre, Trinity College Dublin, Dublin 2,
Ireland}

\author{Alexandre Reily Rocha}
\affiliation{Instituto de Física Teórica, S\~ao Paulo State University, S\~ao Paulo, Brazil}

\author{Caio Lewenkopf}
\affiliation{Instituto de F\'isica, Universidade Federal do Rio de Janeiro, 21941-972 Rio de Janeiro, Brasil}

\date{\today}

\begin{abstract}
Two-dimensional topological insulators (2DTIs) harbor spin-polarized edge states that are topologically protected by time-reversal symmetry against non-magnetic structural disorder. 
However, coupling to magnetic impurities breaks this symmetry, inducing backscattering and destroying perfect quantization. 
While the impact of isolated dilute magnetic impurities is well understood, the transport properties in the presence of dense, disordered ensembles of magnetic moments remain poorly understood. In this work, we develop an analytical framework, supported by extensive numerical simulations, that captures the behavior of edge transport in two-dimensional topological insulators (2DTIs) with a finite concentration of magnetic impurities.
We predict the onset of Anderson localization and uncover an anomalous localization regime characterized by a sub-exponential decay of the conductance, scaling as $\ln {\cal G} \propto -\sqrt{L}$, where $L$ is the system length. 
Furthermore, we demonstrate that the transport exhibits a universal scaling behavior governed solely by the effective impurity concentration. 
Applying our model to Mn-doped HgTe quantum wells, we find excellent agreement with experimental data. 
These findings provide a theoretical foundation for understanding anomalous localization phenomena in magnetically doped topological phases.

\end{abstract}

\maketitle


\section{Introduction} 
\label{sec intro}

Two-dimensional (2D) topological insulators (TIs) have attracted considerable attention in condensed matter physics owing to their topologically protected, spin-polarized edge states. These robust conducting channels, which are immune to backscattering from nonmagnetic disorder, give rise to distinctive transport phenomena and hold significant promise for applications in spintronics and quantum technologies \cite{Culcer2020}.

Prominent examples of topologically protected edge states can be found in Quantum Spin Hall (QSH) systems \cite{Hasan2010, Qi2011}, whose emergence is a result of strong spin-orbit (SO) coupling in inverted-band semiconductor heterostructures. Several 2D materials are predicted to exhibit such behavior \cite{Marrazzo2019}, either intrinsically or through extrinsic mechanisms, through adatom doping \cite{Weeks2011} or proximity-induced effects \cite{Avsar2020}. Far from being confined to the realm of theoretical speculations, experimental realizations have been reported in systems ranging from semiconductor quantum wells \cite{Konig2007, Roth2009, Gusev2011, Knez2014, Du2015} to 2D crystals \cite{Reis2017, Li2017, Fei2017, Wu2018}, and even graphene with adsorbed magnetic clusters \cite{Hatsuda2018}.

A key feature of 2D TIs is the coexistence of time-reversal symmetry and momentum–spin locking, which safeguards edge states against backscattering. As a result, scattering-free edge channels are expected to be robust against structural disorder, yielding a quantized conductance of ${\cal G}_0 = 2e^2/h$ due to the contribution of one spin channel on each edge. However, this picture is modified in the presence of local magnetic moments, which may arise, for example, from magnetic impurities. Such moments locally break time-reversal symmetry and can induce deviations from the quantized conductance. This can lead to a variety of intriguing phenomena, including the Kondo effect \cite{Shamim2021quantized, Maciejko2009} and spin-flip backscattering in edge states \cite{kimme2016backscattering,Novelli2019,jack2020observation,PhysRevB.104.205418,Lima2022b,PEZO2023115358}.

Several approaches have been employed to study their impact on transport. For instance, Maciejko et al. \cite{Maciejko2009} considered a model with Coulomb interactions in the helical edge states coupled to the spin of a charge puddle, predicting different transport regimes depending on the interaction strength and temperature. Similarly, Kurilovich et al. \cite{Kurilovich2017helical} provided a quantitative analysis of conductance deviations in HgTe-based 2DTIs due to a single Mn impurity. Despite these advances \cite{Maciejko2009, Kurilovich2017helical, cheianov2013mesoscopic, altshuler2013localization, vaitkus2022effect, dietl2023quantitative}, a comprehensive and quantitative description of the cumulative effects of a macroscopically large ensemble of magnetic impurities on the conductance of 2DTIs is still lacking. 

In this work, we address this problem by modeling the helical edge states coupled to a collection of point-like magnetic moments. A clear indication that the edge-state topological protection has been lifted is that we do observe the key signatures of Anderson localization for a sufficiently large number of impurities. Remarkably, we also identify the emergence of an anomalous localization regime, characterized by an exponential decay of the conductance with the square root of the device length $L$, in contrast to the standard behavior of the Anderson model \cite{Anderson}.

Interestingly, we observe a scaling law where both the average conductance and its variance depend directly on the number of impurities, suggesting a new universality class for transport in topological insulators with dilute magnetic disorder. We apply our method to the test by computing the conductance of a HgTe-based 2DTI doped with dilute Mn impurities and find that our results are in excellent agreement with experimental observations \cite{Shamim2021quantized}. 

The remainder of the manuscript is organized as follows. In the next section, we present the model employed throughout this study, first considering a single isolated magnetic scatterer and then generalizing the analysis to the case of multiple scatterers. The Results section reveals the coexistence of conventional Anderson localization and an anomalous contribution characterized by a nontrivial scaling with impurity concentration. Following a detailed discussion of the respective fingerprints of these two contributions, we point out experimental evidence that is in agreement with our findings and lends support to our conclusions.

\section{Model for electronic transport}

\begin{figure}[h]
    \centering
    \includegraphics[width=1.\columnwidth]{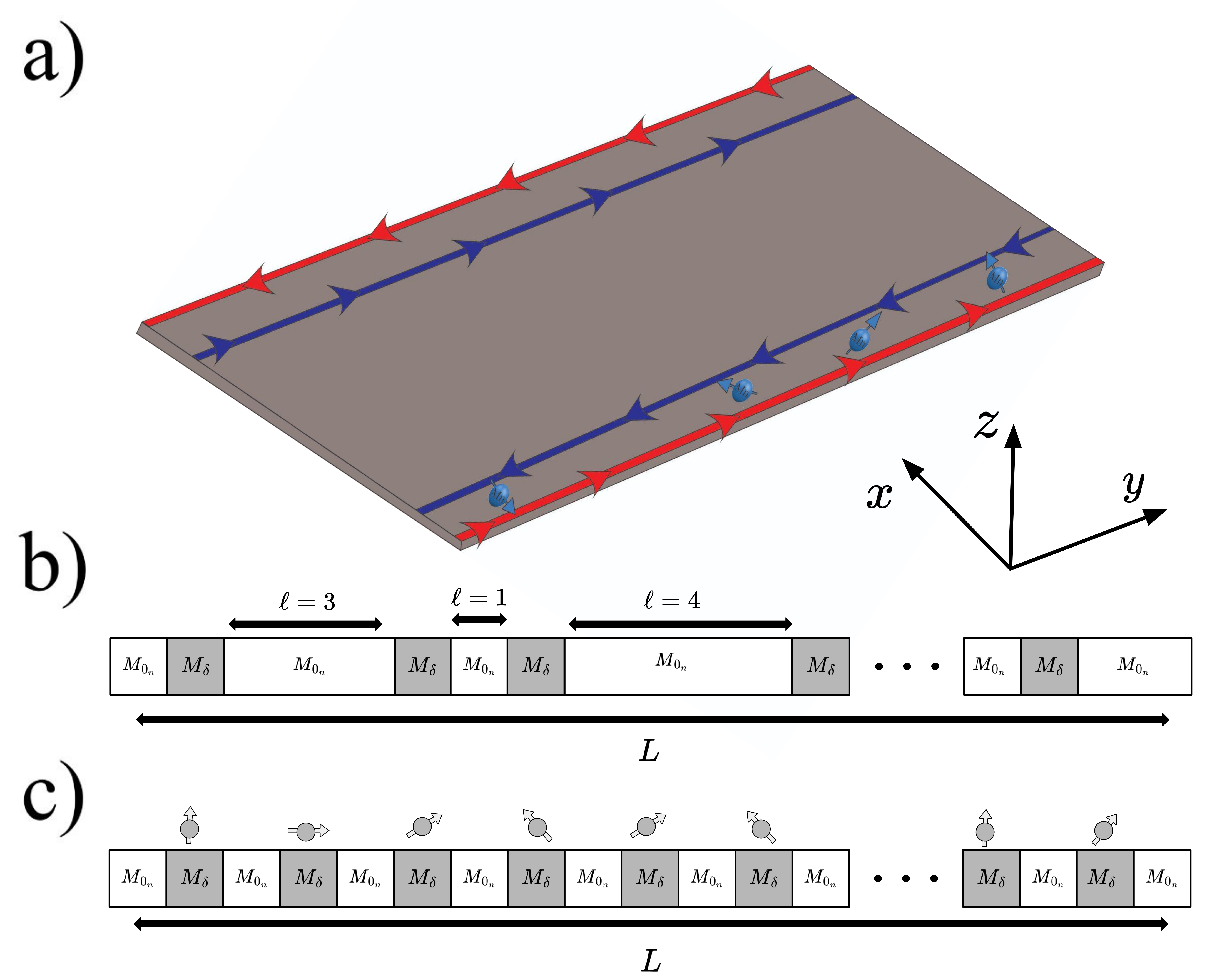}
    \caption{Structure of the quantum wire: 
    a) Shows the edge currents of a topological insulator. Then, we have the lines that can be regarded as the wire, consisting of alternating segments: free propagation regions  \( M_{0_n} \) and scattering regions \( M_\delta \), which represent localized impurities. b) Illustrates a general configuration in which the total length \( L \) can take arbitrary values, and impurities ( \( M_\delta \)) are spatially distributed at random positions. c) Shows a specific realization with spacing \( \ell = 1 \).}
    \label{fig est}
\end{figure}

Our model system consists of a two-dimensional topological insulator (2DTI) described by the Bernevig-Hughes-Zhang (BHZ) model \cite{BHZ2006}, which captures the low-energy band structure of a CdTe/HgTe/CdTe quantum well (QW). 
In the basis of the clean quantum well states, $\{|E,+\rangle, |H,+\rangle, |E,-\rangle, |H,-\rangle\}$, the bulk BHZ Hamiltonian reads
%
\begin{equation} 
\label{Hbhz}
H_{\text{BHZ}}({\bf k})= 
\begin{pmatrix}
H({\bf k}) & 0\\
0 & H^*(-{\bf k})\
\end{pmatrix},
\end{equation}
where the $+ \!\! \uparrow$ block is given by
\begin{equation}
H({\bf k})= \begin{pmatrix}
\varepsilon_k + M_k & A(k_x+ik_y)\\
A(k_x-ik_y) & \varepsilon_k - M_k\
\end{pmatrix}.
\end{equation}
Here, $\varepsilon_k=-Dk^2$, $M_k=M_0-Bk^2$, and $k=\sqrt{k_x^2+k_y^2}$. The parameters $A, B, D,$ and $M_0$ are determined by the quantum well geometry. 
For a typical topological HgTe QW of width $d = 7.0\text{ nm}$, these parameters take well-known values available in the literature \cite{BHZ2006}.

To study the transport properties of the helical edge states (HES), we consider a semi-infinite geometry by introducing a hard-wall confining potential $V(x) = V_0 \Theta(x)$ along the $x$-direction (see Fig. \ref{fig est}), restricting the system to the $x < 0$ half-plane.
In real space, the corresponding Hamiltonian equation reads
\begin{equation} \label{realspaceH}
\left[H_{\text{BHZ}}(-i\partial_x, -i\partial_y) + V(x)\right]\Psi(x,y)=E\Psi(x,y).
\end{equation}

Exploiting the translational invariance along the $y$-direction, we impose periodic boundary conditions such that $k_y$ remains a good quantum number. 
Hence, the solutions of Eq.~\eqref{realspaceH} yield the degenerate helical edge states
\begin{equation} 
\label{Eigenstates}
\Psi^{\uparrow(\downarrow)}_{k_y}(x,y)=\frac{1}{\sqrt{L_y}}e^{ik_y y}\eta_{k_y}^{\uparrow(\downarrow)}(x) \chi^{\uparrow (\downarrow)},\end{equation}
where  $\uparrow,\downarrow$ are the electron spin projections,
$\eta^{\uparrow (\downarrow)}_{k_y}(x)$ is the spatial envelope function, and $\chi^{\uparrow (\downarrow)}$ denotes the four-component spinor. The corresponding linear dispersion relation is given by
\begin{equation}
E^{\uparrow (\downarrow)}_{k_y}=E_0 \pm \hbar v_0 k_y,
\end{equation}
where $E_0=-M_0D/B$ and the Fermi velocity is $v_0=\sqrt{B^2-D^2}|A|/(\hbar |B|)$.

We model the local magnetic moments as classical point-like impurities that couple isotropically to the HES. 
Projecting the full 2D system onto the 1D subspace spanned by the edge states in Eq.~\eqref{Eigenstates}, the effective 1D Hamiltonian for a single edge containing $N$ random magnetic impurities reads
\begin{equation} 
\label{HhfAv}
\mathcal{H} = - i \hbar v_0 \sigma_z \partial_y + \frac{A_0}{2\hbar\rho_\mathrm{imp}} \sum^{N}_{n=1} \bm{\sigma} \cdot {\bf I}_n \delta(y-y_n),
\end{equation}
where $\bm{\sigma}=(\sigma_x, \sigma_y, \sigma_z)$ is the vector of Pauli matrices acting on the helical pseudo-spin space, ${\bf I}_n$ represents the magnetic moment of the $n$-th impurity located at position $y_n$, $A_0$ is the isotropic exchange coupling constant, and $\rho_\mathrm{imp} = N/L$ is the linear impurity density.

We focus on the phase-coherent transport properties of the system. At sufficiently low temperatures, the linear-response edge conductance ${\cal G}$ is given by the Landauer-Büttiker formula ${\cal G}= (e^2/h)T$,
where $T$ is the transmission probability through the disordered region for a single helical channel, and $e^2/h$ is the conductance quantum per spin species. 
For a sample with two counter-propagating edge channels, the total conductance simply becomes ${\cal G} = (2e^2/h) T$.

Before addressing the general scenario, it is instructive to consider the scattering properties of a single magnetic impurity, for which the transmission and reflection coefficients can be written in closed form.  
We treat this single-impurity scattering problem using the transfer-matrix formalism \cite{Markos2008}. 

For a single impurity located at $y_n=0$, the effective 1D Hamiltonian in Eq.~\eqref{HhfAv} simplifies to
\begin{equation}
\label{eq:SingleImp}
\mathcal{H}=
-i\hbar v_0  \sigma_z \partial_y
+ \hbar v_0 \Lambda \delta(y) \left(  
\begin{array}{cc}
\cos \theta & e^{-i\phi}\sin\theta\\
 e^{i\phi}\sin\theta & -\cos\theta\\
\end{array} \right),
\end{equation}
where $\Lambda = ({\hbar v_0})^{-1}A_0|I|/(2\hbar\rho_{\rm imp})$ is the dimensionless exchange coupling strength, and $\theta$ and $\phi$ stand for the polar and azimuthal angles defining the orientation of the magnetic moment ${\bf I}$. 
By integrating Eq.~\eqref{eq:SingleImp} across the $\delta$-function barrier, the transfer matrix $M_\delta$ relating the wavefunctions on either side of the impurity [$\Psi(0^+) = M_\delta \Psi(0^-)$]
is found to be
\begin{equation}
    M_{\delta}=
e^{i\alpha} \left(
\begin{array}{cc}
\cosh \beta  & ie^{-i\phi}\sinh \beta \\
-ie^{i\phi}\sinh \beta  &  \cosh \beta \\
\end{array} \right)  
\end{equation}
where $\alpha=-\Lambda \cos\theta$ and $\beta=-\Lambda \sin\theta$.
The detailed derivation of $M_{\delta}$ is presented in App. \ref{app:TMA}.
As a result, the single-impurity transmission coefficient is independent of the azimuthal phase $\phi$ and reads
\begin{equation}\label{T}
    T_{\delta}=\frac{1}{|[M_{\delta}]_{11}|^2}=\mbox{sech}^2 \beta.
\end{equation}
    
The generalization to the case of multiple impurity scattering consists of taking the product of a chain of transfer matrices,
\begin{equation} \label{M}
M=M_{\delta_N}M_{0_{N-1}}...M_{\delta_2}M_{0_1}M_{\delta_1}    
\end{equation}
where $y_1 < y_2 < \dots < y_{N-1} < y_N$ denote the ordered impurity coordinates along the edge. 
Here, $M_{0_n}$ is the transfer matrix governing free propagation over the spatial segment $\ell_n = y_{n+1} - y_n$ between the $n$-th and $(n+1)$-th magnetic moments, {\it i.e},
\begin{equation}
    M_{0_n}=
\left(
\begin{array}{cc}
e^{ik(y_{n+1}-y_{n})} & 0\\
0 &  e^{-ik(y_{n+1}-y_{n})}\\
\end{array} \right). 
\end{equation}

As illustrated in Fig.~\ref{fig est}b, the transfer matrix of the full scattering chain can be written as an alternating product of single-impurity and free propagation operators. 
To elucidate the algebraic structure of the disorder, it is instructive to evaluate the building block describing scattering from two consecutive impurities separated by a clean propagation segment $\ell_n = y_{n+1} - y_n$ [Fig.~\ref{fig est}(c)]. 
The local composite transfer matrix reads
\begin{widetext}
\begin{multline} \label{product}
M_{\delta_{n+1}} M_{0_n} M_{\delta_n} =
     \cos\!\left[k(y_{n+1}-y_{n})-\frac{\Delta_n^-}{2}\right] e^{i(\alpha_n+\alpha_{n+1})}\left(
\begin{array}{cc}
 e^{i\Delta_n^-/2} \cosh(\beta_n+\beta_{n+1}) & ie^{-i\Delta_n^+/2}\sinh(\beta_n+\beta_{n+1}) \\
-ie^{i\Delta_n^+/2}\sinh(\beta_n+\beta_{n+1}) &  e^{-i\Delta_n^-}\cosh(\beta_n+\beta_{n+1})\\
\end{array} \right) \\
    +i\sin\!\left[k(y_{n+1}-y_{n})-\frac{\Delta_n^-}{2}\right]e^{i(\alpha_{n+1}+\alpha_n)}\left(
\begin{array}{cc}
 e^{i\Delta_n^-/2} \cosh(\beta_n-\beta_{n+1}) & ie^{-i\Delta_n^+/2}\sinh(\beta_n-\beta_{n+1}) \\
ie^{i\Delta_n^+/2}\sinh(\beta_n-\beta_{n+1}) &  e^{-i\Delta_n^-}\cosh(\beta_n-\beta_{n+1})\\
\end{array} \right)
\end{multline}
\end{widetext}
where $\Delta_n^-=\phi_n-\phi_{n+1}$ and $\Delta_n^+=\phi_n+\phi_{n+1}$. 

Equation~\eqref{product} yields a central physical insight: spatial fluctuations in the inter-impurity distance $\ell_n = y_{n+1}-y_n$ enter the phase factors strictly through the combination $k\ell_n - \Delta_n^-/2$. 
Consequently, positional disorder is mathematically redundant with respect to azimuthal phase disorder $\Delta_n^-$.
As we shall see, this redundancy provides the analytical justification for the statistical equivalence observed in Fig.~\ref{fig Twoplot}: spatial disorder may be safely neglected without impacting the macroscopic transmission scaling. 

Accordingly, the full transfer matrix in Eq.~\eqref{M} can be simplified by replacing the random spacing $y_{n+1}-y_n$ with a uniform inter-impurity distance $\ell$. 
Once the simplified full transfer matrix $M$ is constructed, the total transmission coefficient is obtained via $T = 1/\vert{}[M]_{11}\vert{}^2$.


\section{Results}
\label{sec results}

\subsection{Analytical Rationale for Anomalous Localization }

\begin{figure*}[t]
	\centering
	\includegraphics[width=1.\textwidth]{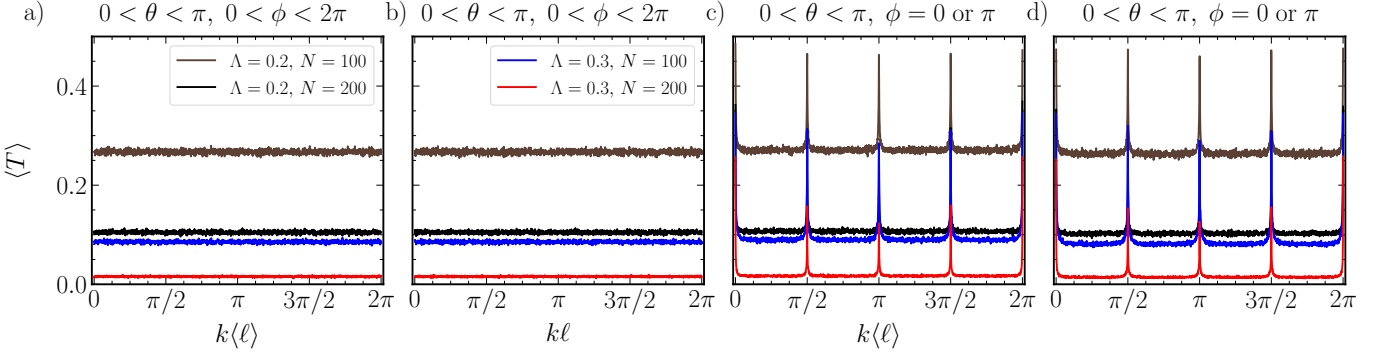}
	\caption{Ensemble-averaged transmission $\langle T \rangle$ as a function of the dimensionless momentum $k\ell$ under different spatial and magnetic disorder configurations.
    (a) Full disorder: independent fluctuations are present in both the polar $\theta_n$ and azimuthal $\phi_n$ angles, as well as the inter-impurity distance $\ell$, with a mean spacing $\langle \ell \rangle = 1$. 
    (b) Full angular disorder with uniform spatial spacing: $\theta_n$ and $\phi_n$ are random, while $\ell_n = 1$ for all sites.
    (c) 
    Polar disorder with constrained azimuthal and spatial parameters: $\theta_n$ is random, while $\phi_n = 0$ and $\ell_n = 1$ are kept fixed. 
    (d) 
    Fully constrained angular and spatial configuration: $\theta_n$ is random, while $\phi_n \in \{0, \pi\}$ and $\ell_n = 1$. 
    Remarkably, $\langle T \rangle$ is statistically equivalent across all configurations, except for panel (d) at the resonant momentum values $k\ell = n\pi/2$ for $n \in \mathbb{N}$. 
    All numerical results were obtained from $2 \times 10^5$ independent disorder realizations.}
	\label{fig Twoplot}
\end{figure*}

To obtain insight into the scaling behavior of the system, we numerically evaluate the ensemble-averaged transmission coefficient $\langle T \rangle$ as a function of the dimensionless momentum $k\ell$ using the transfer matrix method, discussed above. 
The numerical averages are computed over $2 \times 10^5$ independent disorder realizations.

Figure~\ref{fig Twoplot} presents these results for system sizes $N=100$ and $N=200$ at selected exchange coupling strengths, $\Lambda=0.2$ and $\Lambda=0.3$. 
In the most general configuration, represented by case (a) in Fig.~\ref{fig Twoplot}(a), the system accounts for both structural and full angular disorder.
Here, the polar and azimuthal angles are treated as independent random variables uniformly distributed within the intervals $0 < \theta_n < \pi$ and $0 < \phi_n < 2\pi$, respectively. 
Without loss of generality, the average inter-impurity spacing is set to $\langle \ell \rangle = 1$. 
Defining the total system length as 
$L = N(\langle \ell \rangle + 1)$, this configuration establishes a fixed nominal impurity concentration of $\rho_{\rm imp} = N/L = 1/(\langle \ell \rangle + 1)$.

To isolate the microscopic factors driving the localization, we compare this fully disordered case against three increasingly constrained configurations. 
In the second configuration, case (b), spatial fluctuations are suppressed by fixing the inter-impurity distance to the unit value, $\ell_n = \langle \ell \rangle = 1$ (Figure \ref{fig est}c), while maintaining full angular disorder. 
In the third scenario, case (c), spatial disorder is restored, but the azimuthal angle is restricted to a discrete binary distribution where $\phi_n \in \{0, \pi\}$. 
Finally, in case (d), both spatial fluctuations and continuous azimuthal variations are eliminated, leaving the polar angle $\theta_n$ as the sole continuous disordered degree of freedom.

From Fig.~\ref{fig Twoplot}, two immediate conclusions emerge: the average transmission $\langle T \rangle$ decreases monotonically as either the system size $N$ or the coupling constant $\Lambda$ increases. 
A less intuitive observation is that, much like in the single-impurity case, $\langle T \rangle$
remains remarkably independent of the dimensionless momentum $k\ell$ across the bulk of the spectrum. 
Furthermore, comparing panels (a) and (b) reveals that the macroscopic average transmission is statistically invariant under these structural modifications. 
The only distinct feature appears in the fully ordered spatial layout of panels (c) and (d), where sharp Fabry-Pérot-like transmission resonances are evident at the conditions $k\ell = n\pi/2$, with $n$ being an integer.

This striking insensitivity to spatial fluctuations and continuous azimuthal variations indicates that, from the perspective of average transport scaling, the azimuthal phase fluctuations average out while the spatial degrees of freedom effectively decouple. 
It implies that the underlying localization behavior is universally governed by a single effective parameter determined solely by the impurity concentration and the polar angle distribution.

Consequently, one can adopt simplified spatial and angular configurations without any loss of generality. 
This statistical equivalence allows for a drastic simplification of the effective Hamiltonian and its corresponding transport algebra. 
By eliminating the spatial fluctuations, $\ell_n = \ell$, and setting the azimuthal phases to a uniform background, the product of the transfer matrices for successive impurity sites, separated by a clean propagation segment $M_0$, reduces to the compact form
\begin{align} 
\label{Prod}
M_{\delta_{n+1}}M_{0}M_{\delta_n}& = \cos(k\ell){M_{\delta_n+\delta_{n+1}}}+  
\nonumber\\   
+i\sin(k\ell)&e^{i(\alpha_n+\alpha_{n+1})} \nonumber\\
&\!\!\!\!\!\!\!\!\!\!\!\!\!\!\!\!\!\!
 \times\begin{pmatrix}
 \cosh(\beta_n-\beta_{n+1}) & i\sinh(\beta_n- \beta_{n+1}) \\
i\sinh(\beta_n-\beta_{n+1}) &  \cosh(\beta_n-\beta_{n+1})
\end{pmatrix}.
\end{align}

Equation~\eqref{Prod} serves as the building block to evaluate the full chain product in Eq.~\eqref{M}. 
In particular, it allows us to derive an explicit expression for the element $M_{11}$, namely 
\begin{equation} 
\label{M11}
    M_{11}=e^{\left(i\sum_{n}\alpha_n\right)}\sum^{N-1}_{n=0}A_n \sum_{B_n \in {\cal C}_n}\cosh(\beta_N+B_n)
\end{equation}
where $A_n=[\cos(k\ell)]^{N-1-n}[i\sin(k\ell)]^{n}$ and $B_n$ is a given combination of $n$ subtractions and $N-1-n$ additions 
among the remaining set of $N-1$ parameters $\{\beta_1, \dots, \beta_{N-1}\}$. 
Here, ${\cal C}_n$ denotes the set containing all $\binom{N-1}{n}$ distinct sign choices for $B_n$.

As long as $k\ell$ differs from an integer multiple of $\pi/2$,  
the simplified expression in Eq.~\eqref{M11}
is, up to a global phase, statistically
equivalent to the most general 
disorder configuration incorporating arbitrary azimuthal angles
in $\phi_n$. 
The special case of $k\ell = n\pi/2$, with $n$ integer, addressed in Appendix \ref{sec:SU}, leads to a diffusive transport regime.
 
We begin by considering the weak-coupling limit, $\Lambda \rightarrow 0$, expanding $M_{11}$ up to second order in $\Lambda$ 
\begin{equation}
    M_{11}\approx M_0+e^{i\sum_{n}\alpha_n}\sum^{N-1}_{n=0} A_n \sum_{B_n \in C_n}\frac{(\beta_N+B_n)^2}{2} \,\,,
\end{equation}
where $M_0 \equiv M_{11}(\Lambda=0)$ represents the unperturbed ballistic phase factor
\begin{align}
    M_0
    = & \sum^{N-1}_{n=0} A_n \binom{N-1}{n}
\nonumber\\
    = &  \left[\cos(k\ell)+i\sin(k\ell)\right]^{N-1}=e^{ik\ell(N-1)}.
\end{align}
The total transmission in this small-$\Lambda$ regime is given by $T = 1/\vert{}M_{11}\vert{}^2 \approx 1 - 2\operatorname{Re}\{M_0^* (M_{11} - M_0)\}$, which yields
\begin{equation}
\label{T_approx}
    T \approx 1-{\rm Re}\Big\{M_0^*\sum^{N-1}_{n=0}A_n \sum_{B_n \in C_n}(\beta_N+B_0)^2\Big\}.
\end{equation}

Averaging Eq.~\eqref{T_approx} over disorder configurations decouples the phase factors and simplifies the combinatorial sum over $C_n$, leading to the ensemble-averaged transmission up to second order in $\Lambda$
\begin{align} \label{weakCoupl}
\langle T \rangle &\approx 1 - \left\langle \left( \sum_{n=1}^N \beta_n \right)^2 \right\rangle \nonumber \\
&\approx 1 - \frac{\Lambda^2 N}{2}.
\end{align}
The significance of Eq.~\eqref{weakCoupl} lies in the fact that it is valid for the most general case of disorder, both in the polar angles and in the positions of the local magnetic moments, conditions typically encountered in realistic experimental devices.

Most notably, this expansion reveals a powerful scaling property: the dependence of the ensemble-averaged transmission on the parameters $\Lambda$ and $N$ does not depend on them independently, but only through the single composite scaling variable $\Lambda^2 N$, namely
\begin{equation}
	\langle T \rangle(\Lambda, N)\equiv \langle T \rangle(\Lambda^2 N).
\end{equation} 

Additional insight into the general disorder case can be gained by analyzing the structure of Eq.~\eqref{M11}.
We observe that the inner sum spans all possible sign combinations within the arguments of the hyperbolic cosine.
Consequently, there exists a term within this set for which all contributions add constructively, yielding the maximum argument
$\vert{}B\vert{} \equiv \sum_{n=1}^N \vert{}\beta_n\vert{}$. 

Using the momentum invariance of the average transmission across the bulk of the spectrum, we can establish an upper bound for the magnitude $\vert{}M_{11}\vert{}$.
Evaluating Eq.~\eqref{M11} at the representative phase point $k\ell = \pi/4$, where $\cos(k\ell) = \sin(k\ell) = 1/\sqrt{2}$, yields 
\begin{equation}
    M_{11}=\frac{1}{2^{(N-1)/2}}\sum^{N-1}_{n=0} i^{n} \sum_{B_n \in C_n}\cosh(\beta_N+B_n).
\end{equation}
Replacing each hyperbolic cosine term in the summation with its strict upper bound $\cosh(\vert{}B\vert{})$, we obtain
\begin{align} 
\label{Minequality}
    |M_{11}|&<\Bigg|\frac{\cosh(|B|)}{2^{(N-1)/2}}\sum^{N-1}_{n=0} i^{n} \Bigg( \begin{array}{c}
         N-1  \\
         n 
    \end{array}
    \Bigg)\Bigg| 
    \nonumber\\ &= \Bigg|\frac{\cosh(|B|)}{2^{(N-1)/2}}(i+1)^{N-1}\Bigg|=\cosh(|B|).
\end{align}

The inequality in Eq.~\eqref{Minequality} 
allows us to establish a rigorous lower bound for the average transmission,
$\langle T \rangle \ge \langle \operatorname{sech}^2(\vert{}B\vert{}) \rangle$.
In the thermodynamic limit $N \gg 1$, the distribution of the composite parameter $\vert{}B\vert{} = \sum_{n=1}^N \vert{}\beta_n\vert{}$ is governed by the Central Limit Theorem. Consequently, $\vert{}B\vert{}$ obeys a Gaussian probability distribution $P(\vert{}B\vert{})$ characterized by the mean
\begin{equation}
\langle |B| \rangle = N \Lambda \langle |\sin\theta| \rangle,
\end{equation}
and the standard deviation
\begin{equation}
\sigma_{|B|} \equiv \sigma_{\text{SD}}(|B|) = \sqrt{N} \Lambda \sigma_{\text{SD}}(|\sin\theta|).
\end{equation}

By evaluating the ensemble average over this normal distribution, the lower bound for the average transmission takes the integral form
\begin{align} 
\label{eq:inequality}
\langle T \rangle_{N\gg 1} &\ge \int_{0}^{\infty} d|B| \, P(|B|) \operatorname{sech}^2(|B|)  \nonumber \\
&> \frac{2}{\sigma_{|B|} \sqrt{2\pi}} \exp\!\left[ -2\left( \langle|B|\rangle - \sigma_{|B|} \right) \right].
\end{align}
Equation~\eqref{eq:inequality} reveals two competing exponential contributions to the scaling of the lower bound. 
The first term, proportional to $\langle|B|\rangle \propto N$, corresponds to the standard exponential decay associated with conventional Anderson localization, originating from the linear growth of the mean total phase argument $\langle \vert{}B\vert{} \rangle$. 
The second term, proportional to $\sigma_{\vert{}B\vert{}} \propto \sqrt{N}$, represents an anomalous localization effect driven directly by the statistical fluctuations of $\vert{}B\vert{}$ across disorder configurations.


Being an inequality, Eq.\eqref{eq:inequality} 
does not yield an exact expression for $\langle T \rangle$,
but it provides important insights into the underlying transport scaling. One might be tempted to assume a purely exponential decay of the average transmission with system size, $\langle T \rangle \propto e^{-\gamma N}$, 
as in the standard one-dimensional Anderson localization \cite{Anderson}. 
However, as demonstrated in the derivation of Eq.~\eqref{eq:inequality}, $\cosh(\vert{}B\vert{})$ represents the single dominant contribution in the sum of Eq.~\eqref{M11}.
Therefore, even without replacing every term $\cosh(\beta_N + B_n)$ with its maximum value $\cosh(|B|)$, 
the overall ensemble average $\langle T \rangle$ inherits a qualitatively identical functional form
\begin{equation} 
\label{Transmission}
\langle T \rangle(\Lambda^2 N)\propto e^{-(\gamma_1 \Lambda^2 N+ \gamma_2\Lambda \sqrt{N}) } ,   
\end{equation}
where $\gamma_1$ and $\gamma_2$ are dimensionless constants to be determined. 

Equation~\eqref{Transmission} represents the central result of this section: it establishes the coexistence of two distinct localization regimes. 
The first term describes a conventional Anderson-like regime, where the average transmission decays exponentially with the number of scatterers $N$. 
The second term represents an anomalous localization regime in which the transmission decays exponentially with $\sqrt{N}$.

While anomalous localization with $\sqrt{N}$ scaling has been reported in tight-binding models \cite{MudryBrouwerAkira1999,BROUWER2001333,amanatidis2012,molina2020,Inui1994} as a result of off-diagonal site or hopping disorder, the mechanism identified here is fundamentally different. 
In our topological system, this behavior emerges directly from the linear helical dispersion of the edge states combined with the multiplicative statistical fluctuations of the effective coupling parameter $\vert{}B\vert{}$.


By analogy with conventional Anderson localization, one can define a characteristic localization length for the anomalous scaling regime. 
Expressing the total system size as $L = N / \rho_{\text{imp}}$, we propose the scaling ansatz for the ensemble-averaged transmission
\begin{equation} 
\label{Ansatz}
\langle T \rangle(L) \propto \exp\!\left[ -\left( \frac{L}{\xi_S} + \sqrt{\frac{L}{\xi_A}} \right) \right].
\end{equation}
Here, $\xi_S$ and $\xi_A$ represent the standard and anomalous localization lengths, defined respectively as
\begin{equation} \label{LLS}
\xi_S = \frac{1}{\gamma_1 \Lambda^2 \rho_{\text{imp}}}
\quad
\mbox{and}
\quad
\xi_A = \frac{1}{\gamma_2^2 \Lambda^2 \rho_{\text{imp}}}.
\end{equation}


\subsection{Numerical Analysis of Anomalous Localization}

\begin{figure}[hb]
    \centering
    \includegraphics[width=0.892\columnwidth]{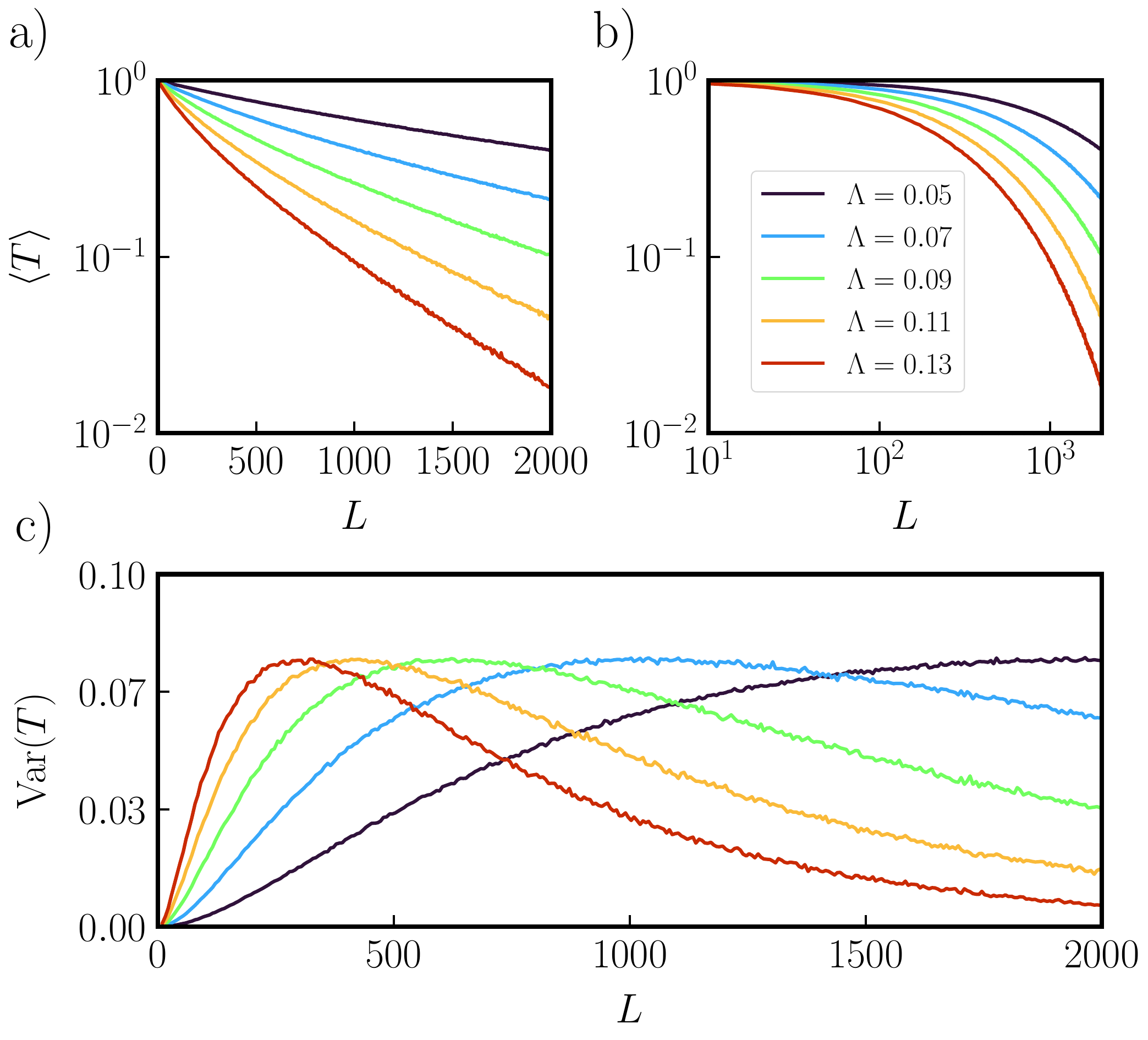}
    \caption{
    Ensemble-averaged transmission $\langle T \rangle$ and transmission variance as a function of system length $L$ for various coupling parameters $\Lambda$. 
    (a) Average transmission on a linear scale. 
    (b) Average transmission on a logarithmic scale. 
    (c) Variance of the transmission, ${\rm Var}(T)$, as a function of $L$. 
    Results in all panels are presented across multiple values of $\Lambda$.}
    \label{fig TandV}
\end{figure}

To fully determine the scaling behavior in the general transport regime, the dimensionless parameters $\gamma_1$ and $\gamma_2$ must be extracted numerically. 
To this end, we perform comprehensive transfer-matrix simulations of the ensemble-averaged transmission $\langle T \rangle$ as a function of system length $L$ across various coupling strengths $\Lambda$.

Figure~\ref{fig TandV} presents these simulation results. 
In Fig.~\ref{fig TandV}(a), $\langle T \rangle$ is plotted as a function of $L$ on a log-linear scale, illustrating the rapid suppression of transmission with increasing size. 
In Fig.~\ref{fig TandV}(b), plots the same on a log-log scale; notably, the data deviate from a simple linear dependence on $L$, indicating that transport cannot be described purely by standard Anderson localization. 
Finally, Fig.~\ref{fig TandV}(c) displays the variance of the transmission, $\operatorname{Var}(T)$, as a function of $L$. 
As the coupling strength $\Lambda$ increases, the relative magnitude of the transmission fluctuations is systematically suppressed, mirroring the self-averaging behavior characteristic of localized states.


\begin{figure}[h]
    \centering
    \includegraphics[width=0.98\columnwidth]{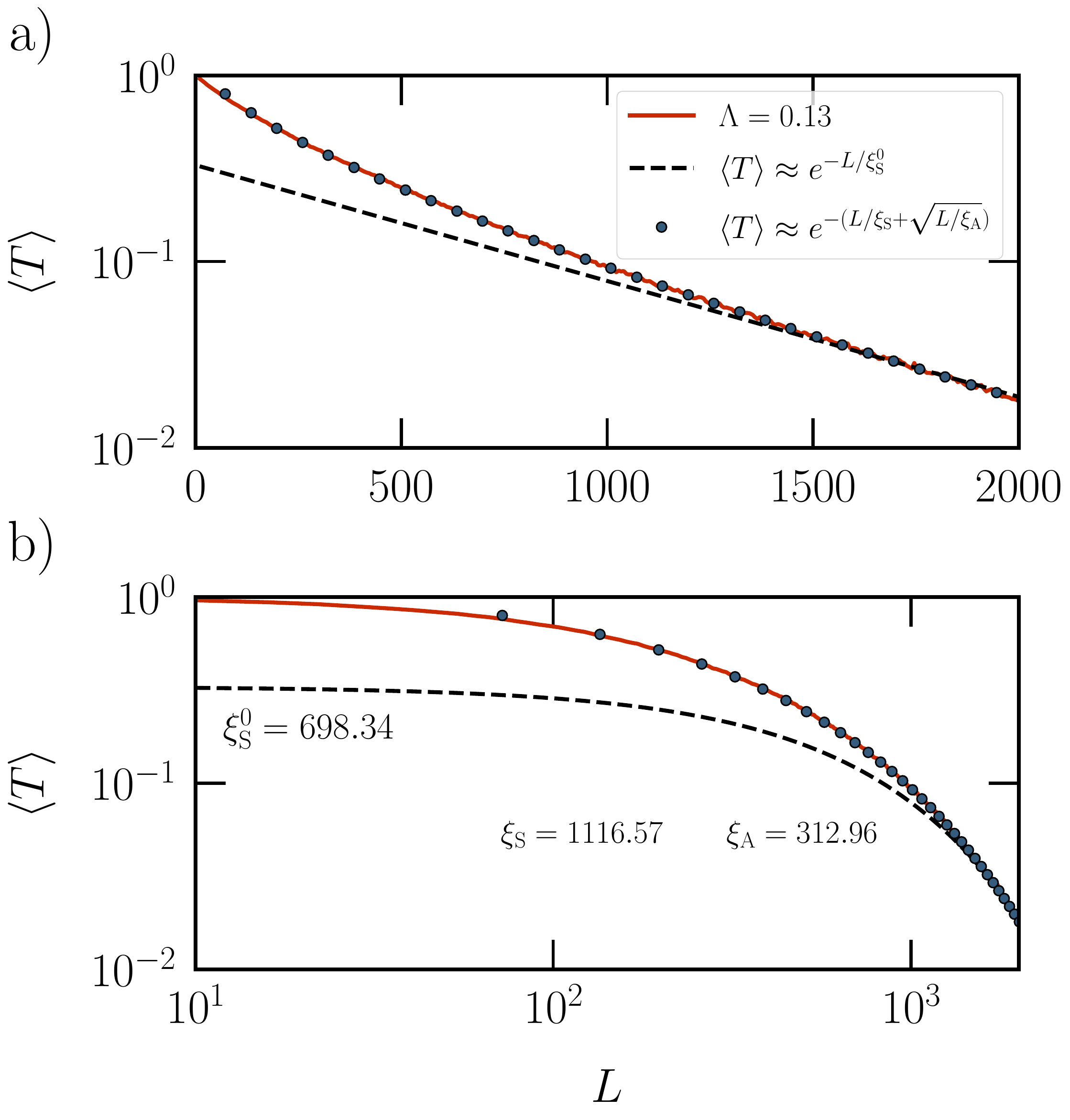}
    \caption{
Average transmission $\langle T \rangle$ as a function of system length $L$ for coupling parameter $\Lambda = 0.13$. 
(a) Transmission plotted on a linear scale. 
(b) Transmission plotted on a logarithmic scale. 
The dots represent numerical simulation data. 
The black dashed line shows a fit assuming conventional Anderson localization, $\langle T \rangle \propto \exp(-L/\xi_S^0)$, which clearly fails to capture the full spatial decay. 
The solid red curve represents a fit incorporating the dual scaling ansatz of Eq.~\eqref{Ansatz}, demonstrating excellent agreement across both short- and long-lengths regimes.
}
    \label{fig TxL}
\end{figure}

For simplicity, in Fig.~\ref{fig TxL} we fix the coupling strength to $\Lambda = 0.13$. Assuming a standard Anderson-like localization, $\langle T \rangle \sim e^{-L/{\xi_{S}^0}}$, a single-parameter exponential fit fails to capture the numerical transmission curve. 
In contrast, 
incorporating the dual scaling form of Eq.~\eqref{Ansatz} yields an exact fit across the entire spatial range, confirming the coexistence of conventional Anderson localization and the emerging anomalous localization regime.

From such numerical fits, both the Anderson localization length $\xi_S$ and the anomalous localization length $\xi_A$ can be quantitatively extracted. 
At small system lengths, $L \ll \xi_S$, the anomalous term dominates transport, governing the decay profile well before standard exponential localization sets in. 
Remarkably, even in the asymptotic regime $L > \xi_S$, where an Anderson-like decay is typically expected to govern the transport, the anomalous scaling contribution remains significant.
Thus, rather than acting as a transient crossover effect, the anomalous term combined with the exponential decay drives the edge state into a robust, mixed-localization regime across all system sizes. 

\begin{figure}[h]
    \centering
    \includegraphics[width=1.\columnwidth]{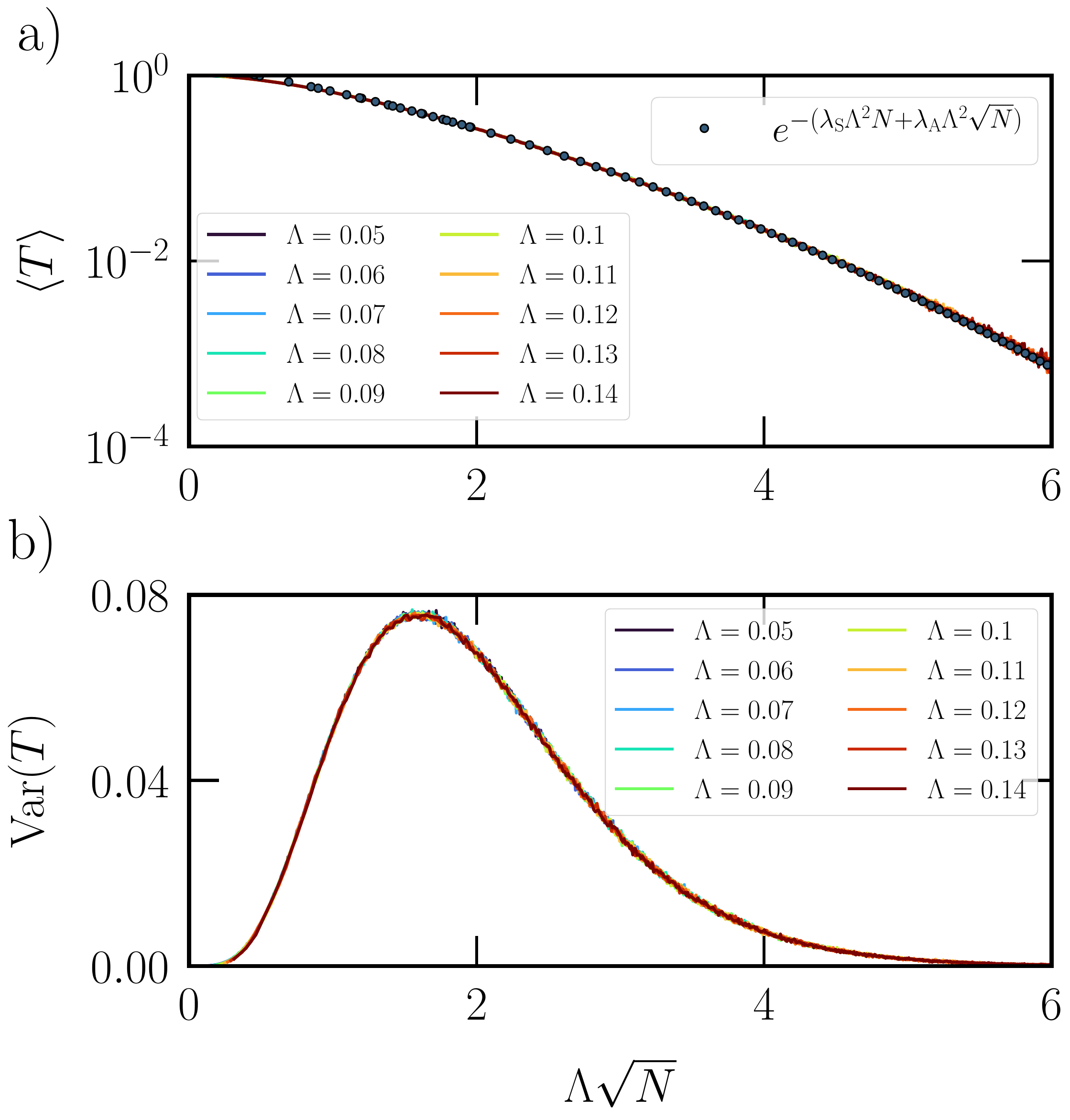}
    \caption{Universal scaling of the transmission with the parameter $\Lambda \sqrt{N}$. 
    (a) Average transmission $\langle T \rangle$ as a function of $\Lambda \sqrt{N}$. 
    (b) Variance of the transmission, ${\rm Var}(T)$, as a function of $\Lambda \sqrt{N}$. 
    In both panels, data collapsing across various coupling strengths $\Lambda$ demonstrates the single-parameter universality driven by both conventional and anomalous localization.}
	\label{fig:Universal}
\end{figure}


In Fig.~\ref{fig:Universal}, when the ensemble-averaged transmission $\langle T \rangle$ is rescaled as a function of $\Lambda \sqrt{N}$, the data collapse onto a single universal curve, fully confirming the analytical scaling predicted by Eq.~\eqref{Transmission}. 
A similar universal collapse is observed for the transmission variance in Fig.~\ref{fig:Universal}(b). 
Analogous to the role played by the localization length ratio $L/\xi$ in standard Anderson localization theory, the combination $\Lambda \sqrt{N}$ serves as the single scaling parameter governing transport across the entire parameter space.

\begin{figure}[!ht]
    \centering 
    \includegraphics[width=1.\columnwidth]{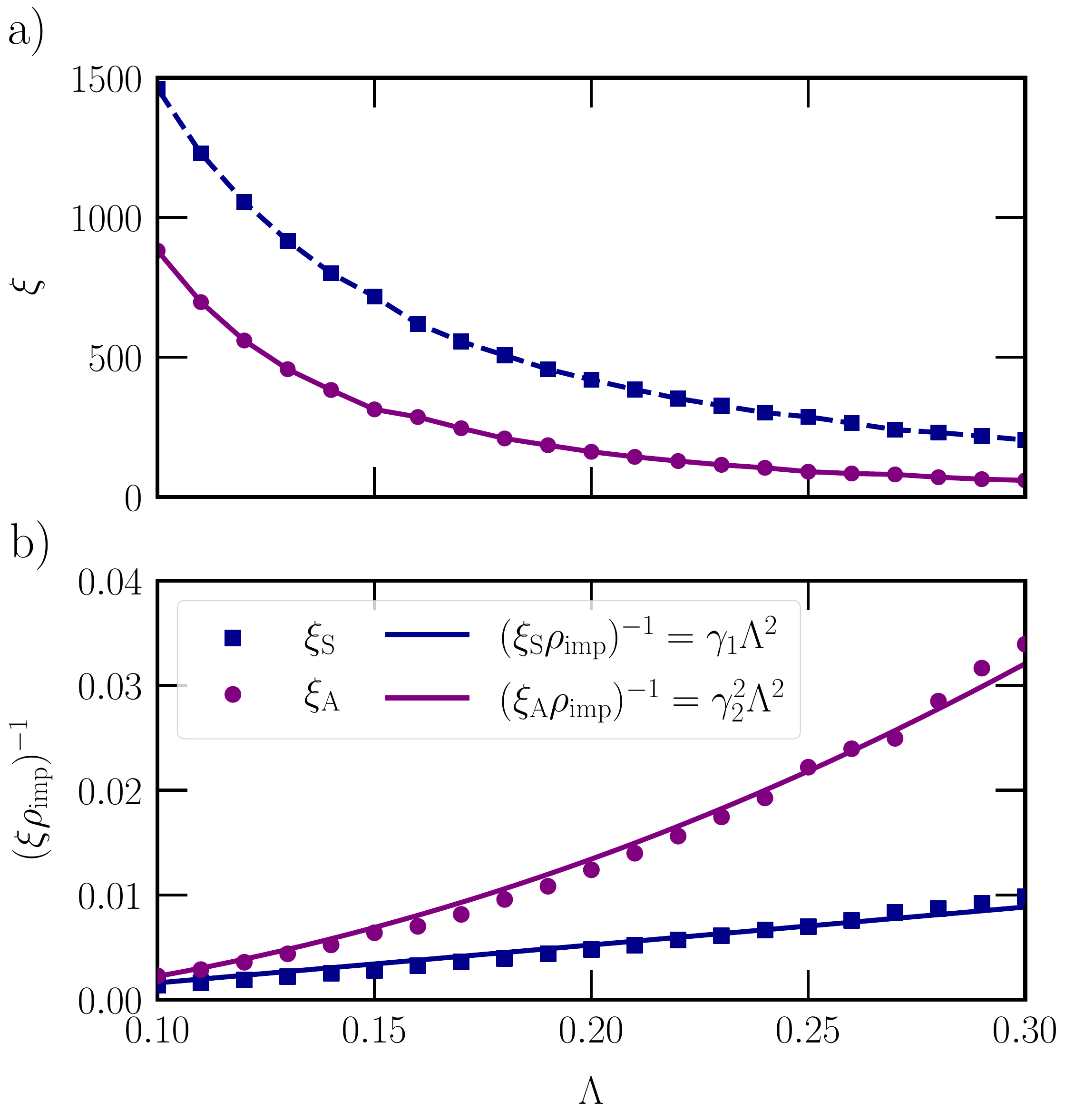}
    \caption{ 
    Scaling coefficients from extracted localization lengths as a function of the coupling parameter $\Lambda$.
    (a) Localization lengths $\xi_S$ and $\xi_A$ versus $\Lambda$.
    (b) Inverse normalized localization lengths $(\xi_S \rho_{\rm imp})^{-1}$ and $(\xi_A \rho_{\rm imp})^{-1}$ as a function of $\Lambda$, confirming the quadratic dependence $\propto \Lambda^2$ predicted by Eq.~\eqref{Transmission}. 
    }
    \label{fig foobar1}
    \label{fig Gamma}
\end{figure}

Finally, to determine the scaling coefficients $\gamma_1$ and $\gamma_2$, we perform a numerical fit of the localization lengths as a function of the coupling parameter $\Lambda$. 
The results are summarized in Fig.~\ref{fig Gamma}.
The upper panel shows the localization lengths displays $\xi_S$ and $\xi_A$ plotted linearly against $\Lambda$. As expected, both localization lengths decrease monotonically with increasing coupling strength $\Lambda$.
As anticipated in the previous paragraph, across the entire investigated parameter range, $\xi_A$ and $\xi_S$ remain within the same order of magnitude, confirming that mixed localization is a pervasive and fundamental feature of the transport. In the lower panel, we present the normalized inverse localization length $(\xi_S \rho_{\rm imp})^{-1}$ and the anomalous coefficient $(\xi_A \rho_{\rm imp})^{-1}$, which exhibit a quadratic dependence on $\Lambda$. Fitting these data (solid curves) allows us to quantitatively extract $\gamma_1=0.114$ and $\gamma_2=0.692$.

\section{Discussion of experimental results}
\label{sec experiments}

The effect of locally breaking time-reversal symmetry in 2D topological insulators has been experimentally studied by Shamim and collaborators using Mn-doped HgTe quantum wells \cite{Shamim2021quantized}. In their work, the authors identify distinct transport regimes governed by temperature: at sufficiently low temperatures  ($T \ll T_K$, where $T_K$ is the Kondo temperature) the average edge conductance approaches the quantized value $\mathcal{G}_0=2e^2/h$ (or $\mathcal{G}_0/2$ per edge). 
This result is interpreted as a manifestation of the Kondo effect, where the localized magnetic moments of the Mn impurities are completely screened by the helical edge electrons. 
As the temperature rises above $T_K$, this screening effect gradually wanes following a power law behavior well described by the experimental data, thus allowing spin-flipping scattering to occur. 
At higher temperatures ($T \approx 15\text{ K}$), a rapid increase in conductance is observed, driven by the thermal activation of bulk carriers, a bulk conduction mechanism that persists even in pristine, impurity-free devices.

There is, however, 
an intermediate temperature region between $4$ and $10\text{ K}$ where an unexpected conductance plateau at approximately $1.48\,e^2/h$ is observed. 
To quantitatively describe the transport properties in this regime, we apply the dual-scaling framework developed above. 
At the modeling level, we consider the exchange interaction between the electron spin and the Mn magnetic moment proposed in Ref.~\cite{Kurilovich2017helical}, where the interaction Hamiltonian takes the form
\begin{equation}
	H_{int}=\frac{1}{2\nu} \mathcal{J}_{ij}I_i\sigma_j\delta(y-y_0)
    \label{eq:Hint}
\end{equation}
where $\mathcal{J}$ is the dimensionless matrix of coupling constants given by 
\begin{equation}
	\mathcal{J}=\frac{2\nu}{\epsilon}e^{-2|x_0|/\epsilon} \left(
	\begin{array}{ccc}
	J_m & 0 & 2J_0\\ 
	0 & J_m & 0 \\
	0 & 0 & J_z \\
	\end{array}.
	\right)
\end{equation}
Here, $\nu=1/(2\pi \hbar v_0)$ is the edge-state density of states per spin species, $\epsilon$ denotes the characteristic width of the edge states, and $x_0$ is the position of the scattering center. 

In the spatial homogeneous limit, Eq.~\eqref{eq:Hint} shares the same operator structure as Eq.~\eqref{HhfAv}, apart from an off-diagonal term proportional to $I_x \sigma_z$ which we neglect in the present analysis. 
For a (001) CdTe/HgTe/CdTe quantum well near the critical well width,
the exchange coupling constants are estimated to be $|J_m| \sim |J_z| \sim |J_0| \sim 0.1~\text{eV}\cdot\text{nm}^2$. 
Using the estimates $\nu \simeq 0.5 ~ \text{eV}^{-1}\cdot \text{nm}^{-1}$ and $\epsilon \simeq 40~\text{nm}$, Kurilovich \textit{et al.} \cite{Kurilovich2017helical} obtained a dimensionless coupling matrix of order $\mathcal{J}_{ij} \sim 10^{-3}$.

Comparing Eq.~\eqref{eq:Hint} to our effective model in Eq.~\eqref{HhfAv} yields a coupling strength of $\Lambda \approx \pi \mathcal{J}_{ij} \approx \pi 10^{-3}$. 
Together with the atomic linear density of $10^7~\text{atoms}/\mu\text{m}$
\cite{Lunde2013}, alongside the $1.2\%$ Mn doping concentration reported in the experiment \cite{Shamim2021quantized}, we obtain an effective impurity density of $\rho_{\text{imp}} \approx 1.2 \times 10^5\,\mu\text{m}^{-1}$.
From these physical parameters, the two localization lengths are quantitatively evaluated as 
\begin{eqnarray} 
\xi_S=\frac{1}{\gamma_1\Lambda^2\rho_S} \approx 28\mu\text{m}  ~,  \\
\xi_A=\frac{1}{\gamma^2_2\Lambda^2\rho_S} \approx  6.6\mu\text{m} ~.
\end{eqnarray}

More importantly, these quantitative localization lengths allow us to directly estimate the expected average conductance of the experimental device. 
The total number of effective impurities along the helical channel is given by $N = \rho_{\text{imp}} L$, where $L = 1.9\,\mu\text{m}$ corresponds to the channel length in the experiment \cite{Shamim2021quantized}. 
This yields a dimensionless scaling parameter of $\Lambda \sqrt{N} \approx 1.05$.
Tracing this value on the universal scaling curve of Fig.~\ref{fig:Universal}(a) yields a single-channel ensemble-averaged transmission of $\langle T \rangle \approx 0.63$. 
Accounting for two counter-propagating helical edge channels, our theory predicts a total edge conductance of $\mathcal{G} \approx 1.26\,e^2/h$. Considering that no free fitting parameters were adjusted, this result shows remarkably good agreement with the experimentally observed plateau value of $1.48\,e^2/h$ \cite{Shamim2021quantized}, providing strong evidence that the anomalous localization regime plays a decisive role in the transport of magnetic topological insulators.


Finally, it is worth pointing out that further transport measurements with modest variations in either the Mn concentration or the device length $L$ will reveal more details about this anomalous regime. This offers a straightforward roadmap for systematic experimental investigations of this non-conventional, mixed-localization phenomenon in magnetic topological insulators.

\section{Conclusions}
\label{sec conclusions}

In summary, we have developed an analytical framework based on the transfer-matrix method to quantitatively describe the localization properties of edge-state transport in two-dimensional topological insulators subject to magnetic impurities disorder.
We identified two distinct disorder cases governed by the rotational dynamics of the magnetic moments: (i) an anisotropic regime, where the polar angle $\theta$ is disordered while the azimuthal angle $\phi$ remains fixed, and (ii) a fully isotropic regime, where spatial fluctuations affect both angles simultaneously.

In the first case, for a large number of impurities $N$, the ensemble-averaged transmission $\langle T \rangle$ becomes virtually momentum-independent, except at critical resonant momenta where the dimensionless Fermi wave vector satisfies $k \ell = m \pi / 2$ ($m \in \mathbb{Z}$). At these discrete resonances, $\langle T \rangle$ deviates from standard exponential suppression, displaying a non-exponential decay with respect to $N$. 


In the second case, which serves as a realistic model for unconstrained disorder realizations, the transmission spectrum remains entirely flat across all momenta. 
However, the spatial decay of $\langle T \rangle$ exhibits a universal anomalous scaling form characterized by an additional $\sqrt{N}$ term in the exponent. 
This feature contrasts sharply with conventional Anderson localization, where spatial decay is strictly proportional to $N$. 
This non-standard decay allows us to define an anomalous localization length $\xi_A$ alongside the conventional localization length $\xi_S$, establishing a dual-length-scale description for transport in helical liquids with magnetic disorder.

These analytical findings are fully corroborated by extensive numerical simulations and our parameter-free agreement with recent experimental data demonstrates the direct physical applicability of this dual-scaling theory to magnetic topological insulator devices.
This result points to the possibility of experimentally studying the phenomena of anomalous localization in electronic systems, a phenomena still unexplored experimentally.


\begin{acknowledgments}
    This work is supported by the Brazilian funding agencies CNPq, FAPERJ, and FAPESP (Project \# 2021/14335-0, 2023/09820-2, 2025/24203-5 and 2025/25217-0).  
    This publication has emanated from research conducted with the financial support of Research Ireland, Grant Number 12/RC/2278\_2, and is co-funded under the European Regional Development Fund under the AMBER award.
    This work used resources of the Centro Nacional de Processamento de Alto Desempenho em São Paulo (CENAPAD-SP) and of the LNCC-Santos Dumont supercomputer.
\end{acknowledgments}

\appendix

\section{Derivation of the single-impurity transfer matrix} 
\label{app:TMA}


In this appendix, we present the explicit derivation of the transfer matrix for a single magnetic scatterer localized on the helical edge state of a two-dimensional topological insulator, as modeled by Eq.~\eqref{eq:SingleImp}.

To properly treat the spatial singularity in the exchange Hamiltonian, we represent the Dirac delta function $\delta(y)$ as the limiting case of a rectangular barrier of width $\varepsilon_1 + \varepsilon_2$:
\begin{equation}
\delta(y) = \begin{cases}
\dfrac{1}{\varepsilon_1 + \varepsilon_2}, & -\varepsilon_2 \le y \le \varepsilon_1, \\[8pt]
0, & y < -\varepsilon_2 \quad \text{or} \quad y > \varepsilon_1,
\end{cases}
\label{AppB:delta_def}
\end{equation}
taking the limit $\varepsilon_1, \varepsilon_2 \to 0^+$ at the end of the calculation.


Inside the scatterer region ($-\varepsilon_2 \le y \le \varepsilon_1$), the stationary Dirac eigenvalue equation $H\Psi(y) = E\Psi(y)$ takes the form
\begin{equation}
\left[ -i \hbar v_0 \sigma_z \partial_y + \frac{\hbar v_0 \Lambda}{\varepsilon_1 + \varepsilon_2} \left( \bm{S} \cdot \bm{\sigma} \right) \right] \Psi(y) = E \Psi(y) ,
\label{AppB:Schrod_inside}
\end{equation}
where $\bm{\sigma} = (\sigma_x, \sigma_y, \sigma_z)$ are the Pauli matrices acting on the pseudospin subspace, and $\bm{S} = (\sin\theta\cos\phi, \sin\theta\sin\phi, \cos\theta)$ represents the orientation of the local exchange field. Explicitly in matrix notation, Eq.~\eqref{AppB:Schrod_inside} reads
\begin{align}
\label{AppB:Schrod_explicit}
\left[ -i \hbar v_0 
\begin{pmatrix} 1 & 0 \\ 0 & -1 \end{pmatrix} 
\partial_y + \right. &
 \\ &\!\!\!\!\!\!\!\!\!\!\!\!\!\!\!\!\!\!\!\!\!\!\!\!\!\!\!\!\!\!\!
\left.\frac{\hbar v_0 \Lambda}{\varepsilon_1+\varepsilon_2} 
\begin{pmatrix} \cos\theta & e^{-i\phi}\sin\theta \\ e^{i\phi}\sin\theta & -\cos\theta 
\end{pmatrix} \right] \Psi(y) = E\Psi(y) .
\nonumber
\end{align}
In the asymptotic free region ($y < -\varepsilon_2$ or $y > \varepsilon_1$), the exchange potential vanishes and the Hamiltonian reduces to the unperturbed Dirac operator
\begin{equation}
-i \hbar v_0 \begin{pmatrix} 1 & 0 \\ 0 & -1 \end{pmatrix} \partial_y \Psi(y) = E\Psi(y) .
\label{AppB:Schrod_outside}
\end{equation}

Taking the limit $\varepsilon_2 \to 0$ in Eq.~\eqref{AppB:Schrod_explicit}, the differential equation for the spinor wave function inside the impurity region simplifies to
\begin{align}
\partial_y \Psi(y) = 
& \left( \frac{iE}{\hbar v_0} B - \frac{i\Lambda}{\varepsilon_1} A \right) \Psi(y) ,
\label{AppB:Derivative_limit}
\end{align}
where $B \equiv \sigma_z$ is the standard Pauli matrix, and the matrix $A$ is defined as
\begin{equation}
A = \begin{pmatrix}
\cos\theta & e^{-i\phi}\sin\theta \\
-e^{i\phi}\sin\theta & \cos\theta
\end{pmatrix} .
\label{AppB:Matrix_A}
\end{equation}


Formally integrating Eq.~\eqref{AppB:Derivative_limit} across the barrier region yields
\begin{equation}
\Psi(y) = \exp\!\left( \frac{i E y}{\hbar v_0} B - \frac{i \Lambda y}{\varepsilon_1} A \right) \Psi(0^-) ,
\label{AppB:Formal_sol}
\end{equation}
where $\Psi(0^-)$ is the wave function immediately preceding the scatterer. Evaluating Eq.~\eqref{AppB:Formal_sol} at the boundary $y = \varepsilon_1$ and subsequently taking the singular spatial limit $\varepsilon_1 \to 0^+$, the energy-dependent term vanishes, yielding the exact boundary condition across the localized scatterer:
\begin{equation}
\Psi(0^+) = e^{-i\Lambda A} \Psi(0^-) .
\label{AppB:Boundary_cond}
\end{equation}


The spatial transfer matrix $M_\delta$ corresponding to a single $\delta$-function exchange scatterer is given by the matrix exponential $M_\delta = e^{-i\Lambda A}$. Since the diagonal and off-diagonal sub-blocks of $A$ commute, we factorize the exponential as
\begin{equation}M_\delta = \exp\!\left[ -i \Lambda \cos\theta \begin{pmatrix} 1 & 0 \\ 0 & 1 \end{pmatrix} \right] 
\exp\!\left[ -i \Lambda \sin\theta \begin{pmatrix} 0 & e^{-i\phi} \\ -e^{i\phi} & 0 \end{pmatrix} \right].
\label{AppB:Transfer_factorized}
\end{equation}
To evaluate these matrix exponentials, we apply standard Pauli spin algebraic identities. 
The scalar phase factor simplifies to
\begin{equation}\exp\!\left[ -i \alpha \begin{pmatrix} 1 & 0 \\ 0 & 1 \end{pmatrix} \right] = e^{-i\alpha} \begin{pmatrix} 1 & 0 \\ 0 & 1 \end{pmatrix} ,
\label{AppB:Phase_factor}
\end{equation}
where $\alpha \equiv \Lambda \cos\theta$. 
For the non-diagonal exchange component, we use the property $(\bm{n} \cdot \bm{\sigma})^2 = \mathbb{I}$, yielding
\begin{align}
\label{AppB:Offdiag_factor}
\exp\!\left[ \beta 
\begin{pmatrix} 
0 & i e^{-i\phi} \\ -i e^{i\phi} & 0 \end{pmatrix} 
\right] = &\\
&\!\!\!\!\!\!\!\!\!\!\!\!\!\!\!\!\!\!\!\!\!\!\!\!\!\!\!\!\!\!\!\!\!\!
\cosh\beta 
\begin{pmatrix} 1 & 0 \\ 0 & 1 \end{pmatrix} + 
\sinh\beta \begin{pmatrix} 0 & i e^{-i\phi} \\ -i e^{i\phi} & 0 \end{pmatrix} ,\nonumber
\end{align}
where $\beta \equiv -\Lambda \sin\theta$.
Combining Eqs.~\eqref{AppB:Phase_factor} and \eqref{AppB:Offdiag_factor}, the single-impurity transfer matrix assumes the desired compact closed-form expression
\begin{equation}
M_\delta = e^{-i\alpha} 
\begin{pmatrix}
\cosh\beta & i e^{-i\phi} \sinh\beta \\
-i e^{i\phi} \sinh\beta & \cosh\beta
\end{pmatrix} ~.
\label{AppB:Transfer_final}
\end{equation}

\section{Transmission at resonant momenta, $k\ell = n \pi/2$ case}
\label{sec:SU}

In the resonant cases where the dimensionless momentum $k\ell$ is an integer multiple of $\pi/2$, the pronounced peaks in the ensemble-averaged transmission can be understood directly from the algebraic structure of the transfer matrix. 
In this regime, the majority of off-diagonal phase terms in $M_{11}$ vanish, thereby minimizing $\vert{}M_{11}\vert{}^2$ and maximizing the overall transmission $T = \vert{}M_{11}\vert{}^{-2}$.

Specifically, $M_{11}$ simplifies to 
\begin{equation}
M_{11} = e^{i\sum_{n}\alpha_n} \cosh(\beta_N + B_0) ~,
\end{equation}
for even multiples of $\pi/2$, and to
\begin{equation}
M_{11} = e^{i\sum_{n}\alpha_n} \cosh(\beta_N + B_{N-1}) ~,
\end{equation}
for odd multiples of $\pi/2$. Consequently, for a fixed azimuthal angle $\phi$, the corresponding transmission probabilities reduce to the compact expressions. 
For $\sin(k\ell) = 0$ one writes
\begin{align}
T_{\phi = \text{const}} = \operatorname{sech}^2\!\left( \beta_1 + \beta_2 + \dots + \beta_N  \right) , \label{AppA:even}
\end{align}
whereas $\cos(k\ell) = 0$ leads to
\begin{align}
     T_{\phi = \text{const}} = \operatorname{sech}^2\!\left( \beta_1 + \beta_2 + \dots - \beta_N \right) . \label{AppA:odd}
\end{align}
%
Remarkably, in this resonant regime, the collective scattering from $N$ impurities maps directly onto the action of a single effective impurity whose coupling strength is determined by the algebraic sum of the individual impurity parameters $\beta_n$.

Further analytical progress follows from the Central Limit Theorem (CLT), which allows us to determine the probability distribution for the parameter sums $B_0 + \beta_N = \sum_{n=1}^N \beta_n$ and $B_{N-1} + \beta_N = \sum_{n=1}^{N-1} \beta_n - \beta_N$. Since both expressions yield statistically equivalent distributions for large $N$, it is sufficient to focus on $\mathcal{B}_N \equiv \sum_{n=1}^N \beta_n$. In the limit $N \gg 1$, the sum $\mathcal{B}_N$ obeys a Gaussian distribution $P(\mathcal{B}_N)$, with mean $\langle\mathcal{B}_N\rangle$ and variance $\sigma_{\mathcal{B}_N}^2$ given respectively by:
\begin{align}
\langle\mathcal{B}_N\rangle &= \left\langle \sum_{n=1}^N \beta_n \right\rangle = -\Lambda N \langle \sin\theta \rangle = 0 , \label{AppA:mean_val} \\
\sigma_{\mathcal{B}_N} &= \sqrt{ \left\langle \left( \sum_{n=1}^N \beta_n \right)^2 \right\rangle } = \Lambda \sqrt{\frac{N}{2}} . \label{AppA:std_dev}
\end{align}


The ensemble-averaged transmission at these resonant momenta can then be evaluated directly by integrating over the Gaussian probability density $P(s)$ for the parameter sum $s \equiv \mathcal{B}_N = \sum_{n=1}^N \beta_n$
\begin{align} 
\label{eq:Anomalous}
    \langle T \rangle_{N\gg 1}&=\int_{-\Lambda N}^{\Lambda N}P(s) \text{ 
sech}^2(s) ds \nonumber\\
&\approx \frac{2}{\sqrt{\Lambda^2 N \pi}}\tanh \Bigg(\frac{\sqrt{\Lambda^2 N}}{4\pi}  \Bigg) \approx \frac{2}{\sqrt{\Lambda^2 N \pi}}
\end{align}
where $P(s) = (\pi N \Lambda^2)^{-1/2} \exp(-s^2 / N \Lambda^2)$ represents the Gaussian distribution of the sum $\sum_{n=1}^N \beta_n$.


We note that the ensemble-averaged transmission decreases monotonically with increasing system size $N$, but does not follow the standard exponential decay predicted by conventional Anderson localization theory. This non-exponential, power-law scaling, has been identified in various tight-binding models, where it is typically tied to the low-energy physics of systems featuring off-diagonal disorder or chiral symmetries \cite{soukoulis1981,Ziman1982}.  In our helical edge state context, this anomalous scaling at resonant momenta emerges directly from the geometric operator structure of the magnetic exchange Hamiltonian.

\begin{figure}[h]
    \centering
    \includegraphics[width=1.0\columnwidth]{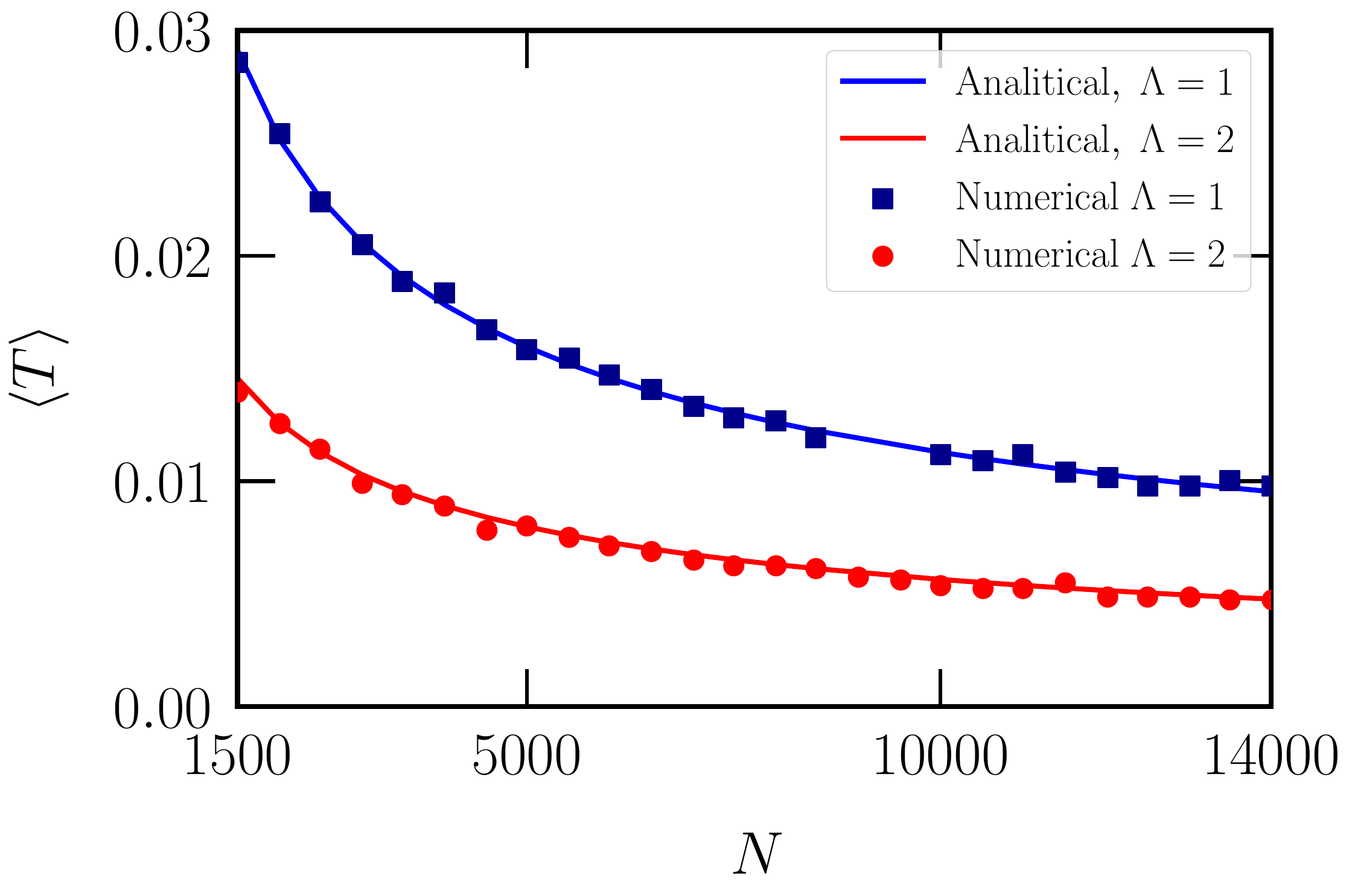}
    \caption{Anomalous decay of the average transmission with the number of scatters. It's possible to observe the good agreement between the analytical prediction and the numerical results. This non-exponential decay is only observed in the absence of disorder in the azimuthal angles of the magnetic moments and for particular values of the dimensionless momenta $k\ell$, when it equates to an integer multiple of $\pi/2$ . }
    \label{fig:AnomalousT}
\end{figure}

This non-exponential power-law decay of $\langle T \rangle$ with respect to scatterer number $N$ is confirmed by the numerical results in Fig.~\ref{fig:AnomalousT}, where the analytical expression in Eq.~\eqref{eq:Anomalous} is compared directly against transfer-matrix simulations for representative coupling strengths $\Lambda = 1.0$ and $\Lambda = 2.0$, and 10,000 realizations.


\bibliography{TIs,localization}

\end{document}